\documentclass[st,twocolumn]{jpsj3}
\usepackage{txfonts}

\usepackage{color}
\usepackage{bm}

\definecolor{migreen}{rgb}{0.0,0.7,0.0}

\title{Quantum many-body solver using artificial neural networks and its applications to strongly correlated electron systems}

\author{Yusuke Nomura$^1$\thanks{yusuke.nomura@tohoku.ac.jp} and Masatoshi Imada$^{2,3}$}
\inst{$^1$Institute for Materials Research, Tohoku University, 2-1-1 Katahira, Aoba-ku, Sendai 980-8577, Japan\\
$^2$Physics Division, Sophia University, Chiyoda-ku, Tokyo 102-8554, Japan\\
$^3$Faculty of Engineering, University of Tokyo, Hongo, Bunkyo-ku, Tokyo 113-8656, Japan} 

\abst{With the evolution of numerical methods, we are now aiming at not only qualitative understanding but also quantitative prediction and design of quantum many-body phenomena. 
As a novel numerical approach, machine learning techniques have been introduced in 2017 to analyze quantum many-body problems. 
Since then, proposed various novel approaches have opened a new era,
in which challenging and fundamental problems in physics can be solved by machine learning methods.
Especially, quantitative and accurate estimates of material-dependent physical properties of strongly correlated matter have now become realized by combining first-principles calculations with highly accurate quantum many-body solvers developed with the help of machine learning methods.
Thus developed quantitative description of electron correlations will constitute a key element of materials science in the next generation.
}

\begin{document}
\maketitle

\section{Introduction}

Machine learning has been applied to physics in various contexts. 
Among them, this review focuses on the 
advances in the application of machine learning techniques to the field of quantum many-body physics.

Understanding quantum many-body systems is a challenge common to many fields, such as electrons in condensed matter physics, nucleons in nuclear physics, and quantum chromodynamics of quarks in particle physics. 
Emergent properties arising from the interplay between quantum nature and many-body nature are at the basis of mechanisms of various intriguing phenomena in the macroscopic world. 

In particular, a grand challenge lies in the strongly correlated systems defined by those where the effects of mutual interaction of particles are essential in the sense that the conventional perturbation expansions from the corresponding noninteracting systems do not work and emergent phenomena show up beyond straightforward mean-field analyses.

Since analytical exact solutions to the quantum many-body problems are not known in most cases of interacting systems, numerical simulations are indispensable for their analyses. 
Remarkable advances in computational science have made it possible to analyze quantum many-body problems with higher precision and on a larger scale than was possible a few decades ago. 
This has not only greatly advanced our understanding of quantum many-body phenomena but also made quantitative predictions possible in some cases. 
In other words, progress in advanced numerical methods can act as an impetus to quantum many-body research. 

In this regard, it is crucially important to develop ``quantum many-body solvers'' that enable us to obtain accurate solutions such as the ground and excited states wave functions as well as thermodynamic and dynamical properties.


In this review, we focus on artificial neural networks developed to variationally obtain accurate ground states and excited states of strongly correlated quantum systems and implemented as tools of quantum many-body solvers.
Such a new trend began in 2017 to analyze strong correlation effects of interacting electrons~\cite{Carleo_2017}, and gradually has also been introduced in nuclear and particle physics~\cite{Adams_2021,Boehnlein_2022,Hayata_2024}. 
Here, we particularly focus on how quantitative descriptions of quantum many-body phenomena in materials (many-electron systems) have been promoted by the implementation of machine learning tools in quantum many-body solvers.
In this context, one of the main topics is applications of solvers that implement artificial neural networks to first principles calculations of strongly correlated electron systems
to elucidate material-dependent properties of materials without adjustable parameters and to understand mechanisms of high-$T_c$ superconductivity (SC) and quantum spin liquid (QSL), which are the subjects in Sec.~\ref{sec:ab_initio}.

This review is organized as follows:
In Sec.~\ref{Sec:MACE}, we introduce a nonempirical framework for strongly correlated materials, consisting of the derivation of low-energy Hamiltonians by first-principles calculations and the analysis of the derived Hamiltonians using low-energy solvers.
Sec.~\ref{Sec:ANN_solver} describes recent progress in the development of a variational low-energy solver employing machine learning techniques.
Several successful applications of neural-network-based variational methods are discussed in Sec.~\ref{Sec:application}.
Finally, we leave several concluding remarks in Sec.~\ref{sec:conclusion}.

\section{Numerical framework for strongly correlated materials}
\label{Sec:MACE}

\subsection{Introduction to many-electron systems and strong correlations}

The motion of electrons in materials is described by the many-body Schr\"odinger equation with the Hamiltonian of
\begin{eqnarray}
{\mathcal H} = 
\sum_{i} \Biggl( - \frac{\hbar^{2}}{2m} \frac{\partial^{2}}{\partial{\bm r}_{i}^{2}} 
-  \sum_{I} \frac{Z_{I}e^{2}}{|{\bm r}_{i}-{\bm R}_{I}|}\Biggr) + \sum_{i<j}  \frac{e^2}{|{\bm r}_{i}-{\bm r}_{j}|}.
\end{eqnarray}
Here, ${\bm r}_i$ (${\bm R}_I$) denotes the position of the $i$th electron ($I$th nucleus) with the charge of $-e$ ($Z_I e$) and the mass of $m$ ($M_I$).
$\hbar$ is the reduced Planck constant.  

In crystalline solids, the lattice potential forms a periodic potential. 
In this case, Bloch's theorem holds, and the energy levels of Bloch states form so-called electronic band structure.
This gives a foundation for understanding the electronic properties of crystalline solids.

The most famous method for calculating band structures is density-functional theory (DFT)~\cite{Hohenberg_Kohn,Kohn_Sham}. 
The DFT is a very useful method; electronic structures are (at least) qualitatively well captured by DFT if the effect of mutual Coulomb interactions of electrons is weak, as in the case of simple metals, band insulators, and semiconductors. 
In those cases, the DFT and the Hartree-Fock approximations capture the essence by taking account of the effects from other electrons as averaged mean fields. 
The many-body effect can be treated by the adiabatic continuation from the noninteracting fermions based on the Fermi liquid theory~\cite{Landau_1956} and perturbative diagrammatic expansions work. 
The machine learning solver has not been well studied in such well-controlled cases and is not so useful actually.

However, there exist certain classes of materials where the single-particle band picture breaks down. 
Most representative example would be a Mott insulator, where the system becomes insulating because of electron localization due to strong Coulomb repulsion. 
The Mott insulator can be realized even with an odd number of electrons in the unit cell. 
However, in this situation, single-particle band theory predicts metallic behavior with partially filled band and can never explain insulating behavior unless spontaneous symmetry breaking such as the antiferromagnetic (AF) order makes the folding of the Brillouin zone leading to the full filling of the split band as in the case of the Slater insulator. 
The genuine Mott insulator emerges even without such symmetry breaking as in the case of QSL.

The description of such strongly correlated materials is a challenge, but at the same time it is a fascinating subject, 
where strong quantum fluctuations and quantum entanglement may reveal unexplored basic concepts and principles of the quantum world. 
They may also lead to novel functional materials.
The machine learning and neural network are expected to help accurate description of such strongly correlated materials and this review focuses on this direction of research. 
In particular, we review recent progress that achieved quantitative understanding of real correlated materials by first principles calculations with the help of machine learning.




\subsection{Effective low-energy Hamiltonians to describe quantum many-body phenomena}

Strong correlation effects usually occur when materials have partially filled bands with narrow bandwidth and have strong local Coulomb interaction. 
Such situation is satisfied when materials have localized atomic orbitals such as $d$ and $f$ orbitals and these orbitals are not closed shell.  
Indeed, various fascinating phenomena arising from strong correlations have been observed in, e.g., transition-metal oxides and heavy-fermion materials, each of which have partially filled $d$ and $f$ orbitals.
Another example is found in molecular solids, where molecular orbitals extending in a unit cell consist of mainly $p$ character near the Fermi level. 
Large inter-molecular distance makes small inter-molecular hopping of the $p$ molecular orbitals, resulting in narrow bandwidth and the strong correlation.

\begin{figure*}[t]
\begin{center}
\includegraphics[width=0.7\textwidth]{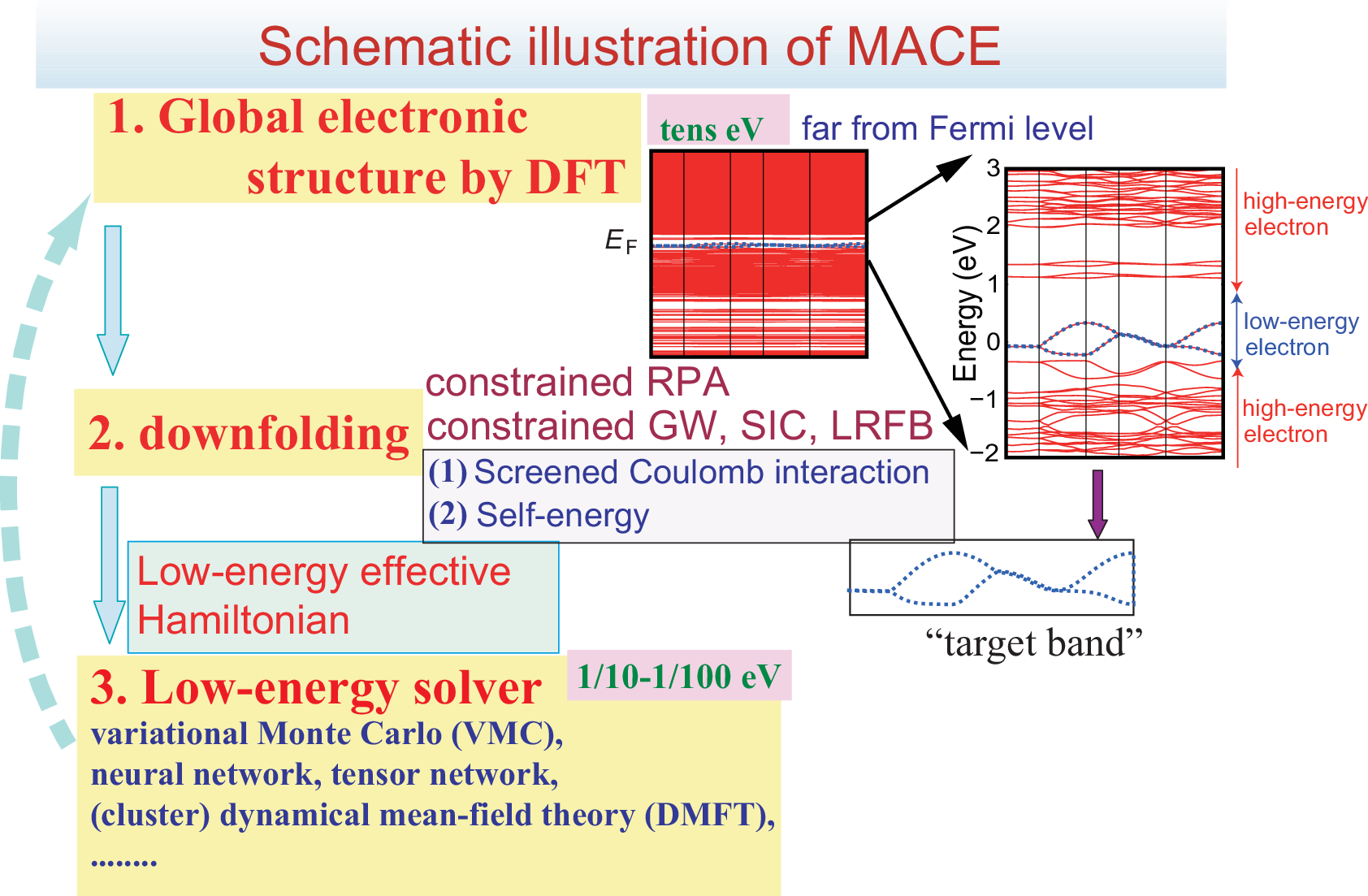}
\caption{Framework of ``multi-energy-scale {\it ab initio} scheme for correlated
electrons (MACE)''~\cite{Imada_2010}. 
}
\label{Fig:MACE}
\end{center}
\end{figure*}

In these strongly correlated materials, partially filled correlated bands emerge on the main stage for low-energy phenomena.
On the other hand, completely occupied/unoccupied bands do not play an essential role. 
In addition, although the bands are dense away from the Fermi level, the bands near the Fermi level is sparse as is required for the strong correlation enabled by poor mutual screening.
Such an energy hierarchical structure in the band structure makes ``multi-energy-scale {\it ab initio} scheme for correlated materials" (MACE) appropriate.
The basic framework of MACE is illustrated in Fig.~\ref{Fig:MACE} and is summarized as:
\begin{enumerate}
    \item Perform DFT band structure calculations and obtain global energy-scale band structure. 
    \item Derive effective low-energy Hamiltonians for correlated subspace including partially filled bands by tracing out high-energy electronic degrees of freedom.
    This procedure is called ``downfolding''. 
    \item Analyze effective low-energy Hamiltonians using accurate quantum many-body solvers.  
\end{enumerate}

In the downfolding, high-energy degrees of freedom play a role in renormalizing one-body hopping parameters and two-body interaction parameters. 
In particular, primary effect of the high-energy band is to screen Coulomb interactions by virtual particle-hole excitations, whose effect can be incorporated within the framework of constrained random phase approximation (cRPA)~\cite{Aryasetiawan_2004}.
Although the RPA is not a well-controlled approximation for the partially filled band of the strongly correlated electrons, the cRPA is, because the low-energy excitations near the Fermi level are excluded, which corresponds to treat as if the system were an insulator~\cite{Imada_2010}.

Several improvements have been made to the cRPA~\cite{Hirayama_2013,Hirayama_2017,Hirayama_2018,Hirayama_2019}: A problem with the cRPA is that the interaction effect partly taken into account in the exchange-correlation energy of the DFT gives double counting of the electron correlation with the correlation effect taken into account in the solution of the low-energy effective Hamiltonian. 
This is resolved by removing the DFT band and replacing it with the GW calculation. 
This is called the constrained GW (cGW) approximation. 
The double counting of the Hartree term in the DFT can also be removed by the self-interaction correction (SIC).
In addition, the relative orbital occupation after solving by the quantum many-body solver in a multi-orbital (multi-band) system can be different from the DFT or GW result, which is the consequence of the ignorance of Madelung potential from the high-energy bands in the quantum many-body solver. 
This can be corrected by level renormalization feedback (LRFB), which shifts the relative chemical potentials of multi-orbitals to recover the DFT or GW result.

Recently, using an idea similar to the cRPA, a scheme for deriving electron-phonon coupling in low-energy Hamiltonians has also been developed~\cite{Nomura_2015_PRB,Falter_note}.
Because it is convenient to define low-energy Hamiltonians in real space, as a basis for low-energy Hamiltonians, localized Wannier orbitals constructed from low-energy subspace are usually employed~\cite{Pizzi_2020,Nakamura_2021}.

When the low-energy subspace consists of a well-isolated single band, 
the effective Hamiltonian is reduced to the following form:
\begin{eqnarray}
\label{eq:Hubbard}
 \mathcal H =  - \sum_{ij} \sum_{\sigma} t_{ij} c_{i\sigma}^\dagger   c_{j \sigma}  + U \sum_i n_{i\uparrow} n_{i \downarrow} + \sum_{i < j} V_{ij}  n_i n_j
\end{eqnarray}
with the hopping parameter $t_{ij}$, onsite Hubbard interaction $U$, and offsite Coulomb interaction $V_{ij}$.
$c_{i\sigma}^\dagger$ ($c_{i\sigma}$) is the electron creation (annihilation) operator at $i$-th site with spin $\sigma=\uparrow$ or $\downarrow$, then $n_{i\sigma}$ and $n_i$ are given by $n_{i\sigma} = c_{i\sigma}^\dagger c_{i\sigma}$ and $n_i =n_{i\uparrow} +n_{i \downarrow} $, respectively.

When the Hubbard $U$ is strong, we get a Mott insulating solution for a half-filled band. 
In this case, an effective model becomes the quantum spin model so-called Heisenberg model
 \begin{eqnarray}
\mathcal H =   \sum_{ i<j  } J_{ij} \   {\bm S}_i \cdot {\bm S}_j 
=  \sum_{ i<j } \frac{J_{ij}}{4}  \left( 
\sigma_i^x \sigma^x_{j} +  \sigma_i^y \sigma^y_{j}  +  \sigma_i^z \sigma^z_{j} 
\right ), 
\end{eqnarray}
strictly speaking, in the strong coupling limit $U/|t_{ij}|\rightarrow \infty$.
Here, $S_i^\alpha = \frac{1}{2} \sigma_i^\alpha$ is the spin-1/2 operator defined at $i$-th site. 
Therefore, 
Heisenberg models can be regarded as specific cases of effective low-energy Hamiltonians.

An important point that we would like to emphasize here is that in these low-energy Hamiltonians, hopping and interaction parameters are no longer tunable or empirical parameters.
Rather, realistic values can be derived only from knowledge about the crystal structure.
In the sense that we do not employ adjustable or empirical parameters, we call these calculations {\it ab initio} or first-principles calculations.

\subsection{Importance of accurate quantum many-body solver}

Once the effective low-energy Hamiltonians are constructed, 
its accurate analysis using quantum many-body solvers such as variational Monte Carlo (VMC) method~\cite{McMillan_1965,Ceperley_1977,Yokoyama_1987_1,Yokoyama_1987_2} and dynamical mean-field theory (DMFT) can lead to quantitative understanding of material properties. 
Indeed, there are various examples where the MACE scheme has succeeded in describing material properties quantitatively, as in the cases of 
cuprates~\cite{Misawa_2016,Hirayama_2019,Schmid_2023}, iron-based superconductors~\cite{Misawa_2011,Misawa_2014,Nomura_2012,Hirayama_2015}, transition-metal oxides~\cite{Okamoto_2014}, ruthenates~\cite{Huebsch_2022}, iridates~\cite{Arita_2012,Yamaji_2014}, fullerides~\cite{Nomura_2015_SciAdv,Kim_2016,Yue_2022,Yue_2023,Witt_arXiv}, and organic materials~\cite{Shinaoka_2012,Misawa_2020,Ido_2022}.

\section{Quantum many-body solver using artificial neural networks}
\label{Sec:ANN_solver}

Since there is no complete and universal numerical method for the analysis of many-body Hamiltonians and each method has pros and cons, it is highly important to make continuous efforts to improve existing methods and develop new low-energy solvers.

Among various types of quantum many-body solvers, VMC is one of the most useful method in analyzing ground-state properties of effective low-energy Hamiltonians. 
In this review, we focus on this VMC-type approach combined with machine learning techniques.

\subsection{VMC method}
\label{Sec:VMC_method}

We consider the problem of calculating the ground state $| \psi_{\rm GS} \rangle$ of effective low-energy Hamiltonian ${\mathcal H}$:
\begin{eqnarray}
    {\mathcal H}| \psi_{\rm GS} \rangle = E_{\rm GS} | \psi_{\rm GS} \rangle, 
\end{eqnarray}
where $E_{\rm GS}$ is the ground-state energy. 
We can expand $| \psi_{\rm GS} \rangle$ using a certain basis $\{ x\}$
\begin{eqnarray}
    | \psi_{\rm GS} \rangle = \sum_{x} \psi_{\rm GS} (x) | x \rangle 
\end{eqnarray}
with the ground-state wave function $\psi_{\rm GS} (x) = \langle x| \psi_{\rm GS} \rangle$ . 
Then, the most straightforward way is to store all the amplitudes $\psi_{\rm GS} (x) \ \forall x$ and compute exactly or approximate $\psi_{\rm GS} (x)$. 
Exact diagonalization and Lanczos method fall under this type of approach. 

However, since the dimension of the Hilbert space grows exponentially with respect to the system size, such an approach is limited to small system sizes (several tens sites).
Instead, VMC approximates the ground state by some parameterized function $\psi_\theta (x)$ (variational wave function) with the set of variational parameters $\theta$.
We only store variational parameters whose number grows algebraically with respect to the system size, and the wave function amplitude is computed by taking a certain $x$ as input. 
This leads to an exponential reduction of required memory size.

One can show that the energy expectation value $E_\theta$ of the variational state $|\psi_\theta \rangle$ cannot be lower than the true ground-state energy
\begin{eqnarray}
    E_\theta = \frac{\langle \psi_\theta | {\mathcal H} | \psi_\theta \rangle }{ \langle \psi_\theta | \psi_\theta \rangle} \geq E_{\rm GS}.
\end{eqnarray}
This variational principle gives a guide for the optimization of the variational state; 
we usually optimize variational parameters $\theta$ to minimize $E_\theta$.

Therefore, in the VMC calculation, one has to estimate $E_\theta$ and the gradient of the energy with respect to variational parameters $\frac{\partial E_\theta}{\partial \theta_k}$ to find the direction to lower $E_{\theta}$ in the $\theta$ space. 
For this purpose, we employ Monte Carlo method. 
To see how it works, we recast the energy expectation value as 
\begin{eqnarray}
 E_\theta = \frac{ \sum_{ x, x'}  \psi_\theta^* (x) 
 \mathcal{H}_{ x x'}  \psi_\theta (x') } { \sum_x \left | \psi_\theta (x ) \right |^2 } 
= \sum_x P_\theta(x) E^{\rm loc}_\theta (x) 
\end{eqnarray}
with 
\begin{eqnarray}
P_\theta (x) = 
\frac{ \left | \psi_\theta (x) \right |^2 } { \sum_x \left | \psi_\theta (x) \right |^2 }, \quad 
E^{\rm loc}_\theta(x) =  \sum_{x'}  \mathcal{H}_{x x'}  \frac{\psi_\theta (x')}{\psi_\theta(x)}.
\end{eqnarray}
As for the gradient vector ${\bm g}_\theta$ whose element is given by $\bigl({\bm g}_\theta \bigr )_k = \frac{\partial E_\theta}{\partial \theta_k}$, 
by defining a vector $ \Bigl({\bm O}^{\rm loc}_\theta (x ) \Bigr)_k = \frac{\partial \log \psi_\theta(x) }{\partial \theta_k}$, we can derive the following formula
\begin{align}
  {\bm g}_\theta  \! =\!   2 {\rm Re} \ \! \biggl( \sum_x P_{\! \theta}(x)  E^{\rm loc}_\theta(x)  {\bm O}_\theta^{\rm loc} (x) \biggr)  - 2  E_\theta  {\rm Re } \biggl(  \sum_x P_{\! \theta}(x) {\bm O}_\theta^{\rm loc} (x)\biggr). 
  \nonumber \\
 \label{Eq.gradient}
\end{align}
Keeping in mind $\sum_x P_\theta(x)=1$, we see that $E_\theta$ and ${\bm g}_\theta$ can be computed from the information of Monte Carlo averages of $E^{\rm loc}_\theta(x)$, ${\bm O}_\theta^{\rm loc} (x)$,
and 
$E^{\rm loc}_\theta(x)  {\bm O}_\theta^{\rm loc} (x)$, where the sampling is performed using the weight proportional to $| \psi_\theta (x)  |^2$.

An important point here is that because the weight $| \psi_\theta (x)  |^2$ is nonnegative, the VMC method has the advantage of avoiding the sign problem in Monte Carlo sampling.
Also, when the Hamiltonian matrix is sparse (i.e., for a given $x$, ${\mathcal H}_{x,x'} = \langle x | {\mathcal H} | x' \rangle $ becomes nonzero 
only for the number of $x'$ linearly scaled with the system size), $E^{\rm loc}_\theta(x)$ can be efficiently computed.
Since this condition is usually satisfied for low-energy Hamiltonians in condensed-matter physics, the total computational cost of the VMC is reduced to a polynomial time with respect to the system size.

The minimization of the parameters is performed until the gradient $\frac{\partial E_\theta}{\partial \theta_k}$ becomes negligibly small. 
At this point, we notice a striking similarity between the VMC task and the machine learning task. 
If we consider the variational energy $E_\theta$ as a cost function and construct $\psi_\theta(x)$ by artificial neural networks, then the task of finding ground state can be considered as a kind of machine learning task in which we optimize nonlinear function (artificial neural networks) using the cost function (energy) in a high-dimensional parameter space~\cite{Melko_2019}.
This point is discussed in more detail later. 

Because both VMC and machine learning have the task of optimizing parameters using gradient, they share some common optimization methods.
The simplest optimization method is called stochastic gradient descent (SGD)
\begin{eqnarray}
 \theta_k^{(t+1)} =  \theta_k^{(t)} - \eta g_k^{(t)} 
\label{Eq.SGD}
\end{eqnarray}
at $t$-th step optimization. $\eta$ is a small positive number (called the learning rate in the machine learning community).  
Nowadays, there exist various types of extensions of SGD such as Adam (adaptive moment estimation)~\cite{Kingma_2014}. 

In VMC, we usually employ a more sophisticated optimization technique called stochastic reconfiguration (SR) method~\cite{Sorella_2001} (natural gradient method in the machine learning community~\cite{Amari_1996,Amari_1998}). 
One can show that the optimization of variational wave function using the SR method reproduces the imaginary-time evolution of quantum states as accurately as possible within the representability of variational ansatz.  
Because the long imaginary-time evolution of quantum state $e^{-\tau {\mathcal H}} | \psi^{(0)} \rangle$ ($\tau \rightarrow \infty$) converges to the ground state as long as the initial state is not orthogonal to the ground state, 
the optimization is greatly stabilized by the SR method.
As for the practical detail of the SR method, see, e.g., Refs.~\citen{Misawa_2019,Nomura_2023}.


\subsection{Artificial neural network for variational ansatz}
\label{Sec:VMC_NN}

\subsubsection{Motivation to introduce artificial neural networks}

In VMC, the accuracy of the ground-state approximation highly depends on the form of the variational wave function.  
Therefore, it is very important to construct a good variational ansatz. 
In the conventional VMC method, the variational wave function is constructed based on some physical insight. 
The variational ans\"{a}tze based on brilliant intuitions have led to a revolution in our understanding of quantum phenomena, for example, Bardeen-Cooper-Schrieffer (BCS) wave function for conventional SC~\cite{Bardeen_1957}, 
resonating valence bond (RVB) wave function for QSL~\cite{Anderson_1973}, and Laughlin wave function for fractional quantum Hall effect~\cite{Laughlin_1983}.

However, in situations such as strongly correlated systems, where various quantum states (each corresponding to a different phase in the thermodynamic limit) compete on small energy scales, the conventional approach that narrows down the subspace for ground-state search may bias the solution and result in qualitatively incorrect predictions.  
Quantitative improvement of accuracy is also difficult because it is highly nontrivial to identify what kind of variational parameters should be added for improvement. 

\begin{figure}[t]
\begin{center}
\includegraphics[width=0.45\textwidth]{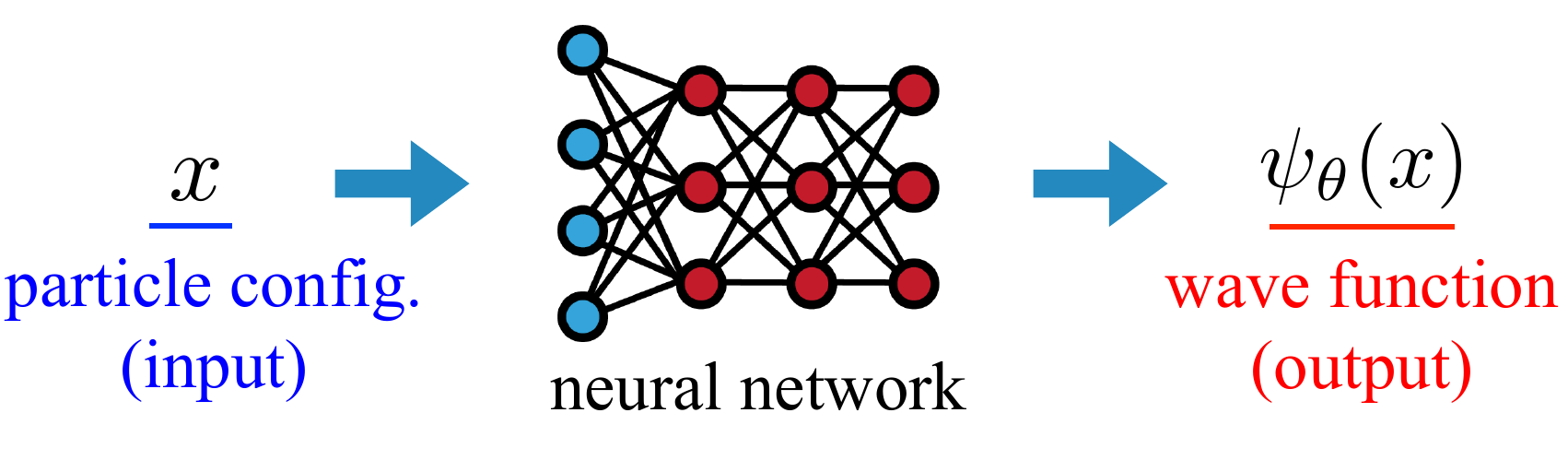}
\caption{Variational wave function constructed from artificial neural networks.
In the conventional variational ansatz, we determine the functional form of $\psi_\theta(x)$ based on some physical insight. 
In the case of neural-network wave function, the functional form is determined based on data.}
\label{Fig:ANN_WF}
\end{center}
\end{figure}

\begin{figure}[t]
\begin{center}
\includegraphics[width=0.49\textwidth]{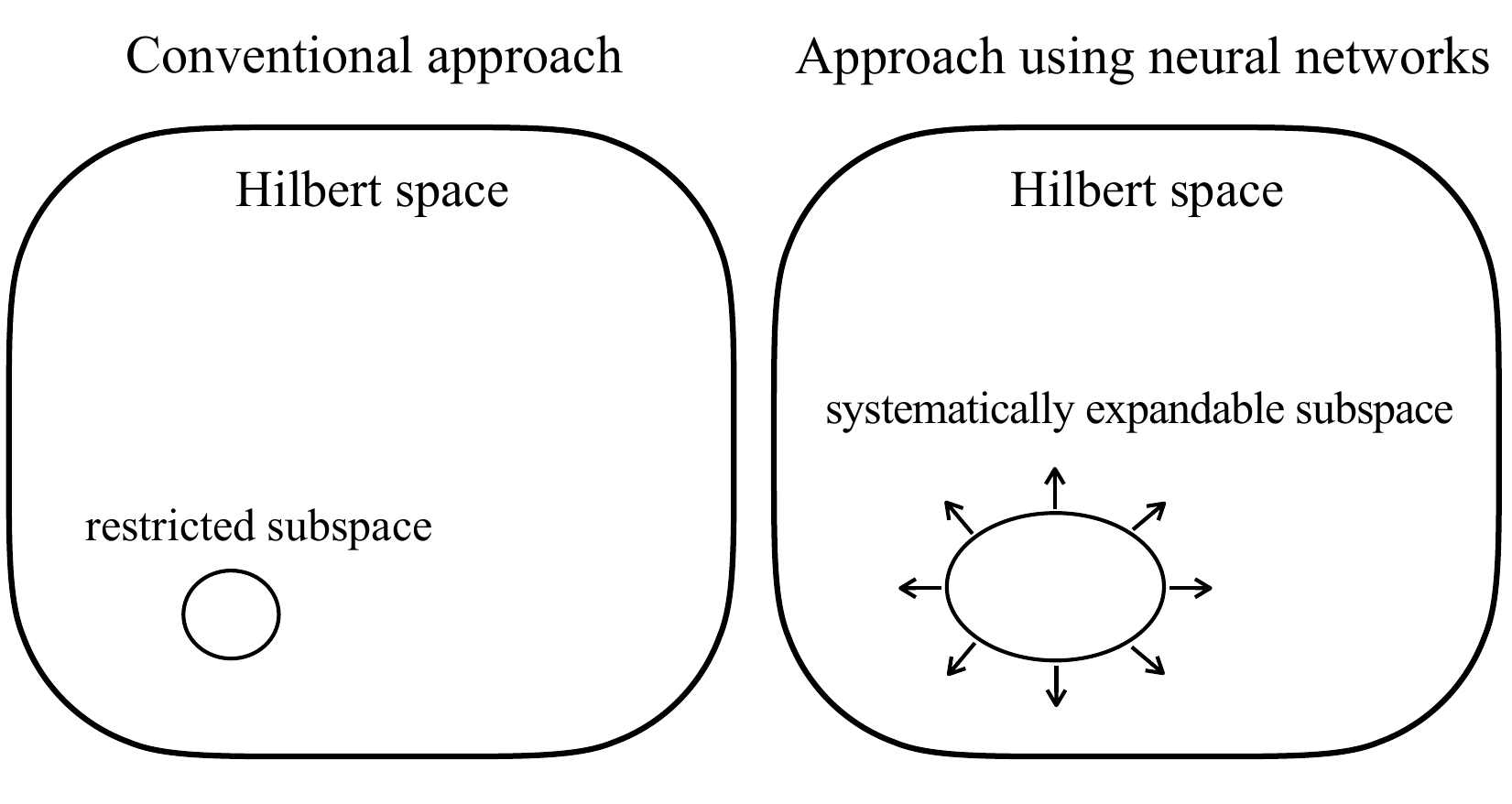}
\caption{While the conventional VMC method narrows down the subspace for the ground-state search, a data science approach using artificial neural networks searches the ground state in systematically expandable subspace.}
\label{Fig:Hilbert}
\end{center}
\end{figure}

Given this situation, it is beneficial to introduce a data science approach and construct a variational wave function by means of an artificial neural network (Fig.~\ref{Fig:ANN_WF}).
Expected benefits are, for example, as follows:
\begin{enumerate}
    \item It helps to mitigate unnecessary biases in searching the true solution. 
    \item Artificial neural networks are known as universal approximators and can approximate any function with arbitrary precision in the limit of increasing network size~\cite{Hornik_1989}.
    Thus, in principle, it is possible to construct variational wave functions that can be systematically improved (Fig.~\ref{Fig:Hilbert}).
\end{enumerate}
In the approach using artificial neural networks, the number of variational parameters can be much larger than that of the conventional approach. 
Therefore, recent advances in numerical techniques and computational resources have made such an approach possible.
A common trend toward less biased variational calculations can be seen, e.g., the extension of VMC method to many-variable variational Monte Carlo (mVMC)~\cite{Tahara_2008,Misawa_2019} and tensor network methods~\cite{Verstraete_2008,Orus_2014}, including the density matrix renormalization group (DMRG)~\cite{White_1992,White_1993} using matrix product states.

\subsubsection{First application of artificial neural networks as a variational ansatz}

Artificial neural networks generate output ${\bm y}$ from input ${\bm x}$ by performing a series of linear and nonlinear operations (${\mathcal L}$ and ${\mathcal N}$, respectively):
\begin{eqnarray}
{\bm y} = \mathcal{N}_{L} \circ \mathcal{L}_{L}\circ \cdots \circ \mathcal{N}_1 \circ \mathcal{L}_1({\bm x}).
\label{eq:ffnn_output}
\end{eqnarray}
In the context of VMC, the input ${\bm x}$ is usually a real-space configuration of particles, and the output ${\bm y}$ is the wave function amplitude $\psi_\theta(x)$ (Fig.~\ref{Fig:ANN_WF}).

In 2017, Carleo and Troyer pioneered such a VMC-type application using the restricted Boltzmann machine (RBM) and applied it to spin-1/2 quantum spin Hamiltonians~\cite{Carleo_2017}.
The RBM is a shallow neural network with one hidden layer using the hyperbolic cosine function as the nonlinear transformation. 
By identifying the input ${\bm x}$ with the spin configuration $\sigma = (\sigma_1, \ldots, \sigma_N)$ ($N$ is the number of quantum spins), the RBM wave function is written as 
\begin{eqnarray}
    \psi_\theta (\sigma) = \prod_{k=1}^M {\rm cosh} \Biggl( b_k + \sum_{i=1}^{N} W_{ik} \sigma_i \Biggr),
\end{eqnarray}
where $b_k$ and $W_{ik}$ are the variational parameters to be optimized and we omit the normalization factor for simplicity. 
The quantum wave function can take complex amplitude; in such cases, we need to employ complex variational parameters (or represent the phase and absolute value of the wave function separately). 
It allows an important flexibility, where the nodes of the wave function can be tuned and optimized to mimic the nodal structure of the exact ground state.
In the RBM ansatz, the number of hidden units $M$ is the control parameter for the representability. 

When introducing a novel variational ansatz, it is important to perform benchmark calculations to check the reliability. 
For this reason, benchmarking for accuracy validation is a main result in Ref.~\citen{Carleo_2017}, and it has been demonstrated that the RBM can represent ground states of frustration-free quantum spin Hamiltonians with high accuracy.

\subsubsection{Extension to challenging low-energy Hamiltonians}
\label{sec:extension_challenging}

Although Ref.~\citen{Carleo_2017} shows a striking result that even one of the simplest artificial neural networks (RBM) can be a powerful variational ansatz, its application is limited to frustration-free quantum spin systems having no competing interactions (frustration effect in quantum spin systems will be discussed in detail in Sec.~\ref{sec:J1J2}).
However, for frustration-free quantum spin systems, the quantum Monte Carlo method based on, e.g., path-integral formalism can be applied without negative sign problems and can give more reliable results than variational approaches.

However, in general quantum many-body systems, the quantum Monte Carlo method suffers from the negative sign problem, originating from negative wave function amplitude or, more generally, complex amplitude.
Even in such cases, the variational methods do not suffer from the sign problem in the Monte Carlo sampling part (see Sec.~\ref{Sec:VMC_method}).
Thus, the main target of variational methods should be such challenging systems like frustrated quantum spin systems and itinerant strongly correlated electron systems.

In fact, there has been extensive effort to extend the VMC approach using artificial neural networks to challenging quantum many-body systems such as frustrated quantum spin systems~\cite{Cai_2018,Liang_2018,Choo_2019,Ferrari_2019,Westerhout_2020,Szabo_2020,Nomura_2021_JPCM,Nomura_2021_PRX,Astrakhantsev_2021,M_Li_2022,Rath_2022,Roth_2023,Viteritti_2022,Viteritti_2023,Viteritti_arXiv,Reh_2023,Chen_2024}, itinerant boson systems~\cite{Saito_2017,Saito_2018},
fermion systems~\cite{Cai_2018,Nomura_2017,Luo_2019,Han_2019,Choo_2020,Pfau_2020,Hermann_2020,Stokes_2020,Yoshioka_2021,Inui_2021,Moreno_2022,Cassella_2023,Imada_2024},
and fermion-boson coupled systems~\cite{Nomura_2020}.
Among them, here, we introduce RBM+PP wave function~\cite{Nomura_2017}, which has realized beneficial applications in the elucidation of physics.
By taking the Hubbard model, one of the representative electronic low-energy Hamiltonians, as an example, we describe the form of the RBM+PP wave function. 
We take the Fock state in real space as the basis
$ | x \rangle = | n_{1\uparrow}, 
\ldots,  n_{N\uparrow}, n_{1\downarrow}, \ldots, n_{N\downarrow} \rangle$, and define the input vector for the artificial neural network as 
\begin{eqnarray}
 {\bm x} = (2n_{1\uparrow}-1 , 
\ldots,  2n_{N\uparrow}-1, 2n_{1\downarrow}-1, \ldots, 2n_{N\downarrow}-1 ),
\end{eqnarray}
where ${\bm x}$ has $2N$ components and each element takes the value of $\pm 1$, 
representable as pseudo-Ising-spin variables.

\begin{figure}[t]
\begin{center}
\includegraphics[width=0.49\textwidth]{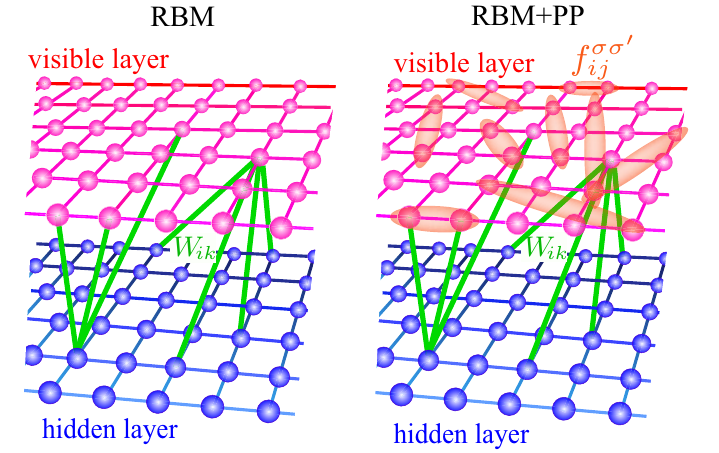}
\caption{
Comparison between RBM and RBM+PP wave functions. 
In the RBM, quantum many-body correlations are incorporated through the coupling to the hidden layer. 
On the other hand, RBM+PP incorporates important correlations in the PP part and helps the RBM part learn the ground state. 
For visibility, only a limited number of $W_{ik}$ and $f_{ij}^{\sigma\sigma'}$ parameters are shown.
Adapted with permission from Ref.~\citen{Nomura_2017}. Copyright 2017 by the American Physical Society.
}
\label{Fig:RBM_PP}
\end{center}
\end{figure}

The RBM+PP wave function is defined as a product of the RBM part and PP part.
The RBM part is given by 
\begin{eqnarray}
    \phi_{\rm RBM} (x) = \prod_{k=1}^{M} {\rm cosh} \Biggl( b_k + \sum_{i=1}^{2N} W_{ik} x_i \Biggr).
\end{eqnarray}
The pair-product (PP) part is given by 
\begin{eqnarray}
    | \phi_{\rm PP} \rangle = \Biggl (  \sum_{i, j=1}^{N}  \sum_{\sigma, \sigma'=\uparrow,\downarrow } f_{i  j}^ {\sigma \sigma'} c^{\dagger}_{i \sigma} c^{\dagger}_{j \sigma'}   \Biggr )^{N_{\rm e}/2}|0\rangle,
\label{eq_PP_state}
\end{eqnarray}
where $N_{\rm e}$ is the number of electrons, $f_{i  j}^ {\sigma \sigma'}$ are variational parameters.
The value of $\phi_{\rm PP}(x)=\langle x | \phi_{\rm PP}\rangle$ is expressed as the Pfaffian of a skew-symmetric matrix whose elements depend on $x$ and $f_{i  j}^ {\sigma \sigma'}$~\cite{Tahara_2008}.

The advantage of combining the RBM wave function with the PP wave function is as follows:
\begin{enumerate}
    \item Although the RBM part is symmetric with respect to the exchange of particles, because the PP wave function is anti-symmetric, the total RBM+PP wave function becomes anti-symmetric. Then, the RBM+PP wave function can be applied to electron (fermion) systems. 
    \item The PP part takes into account important correlations and helps the RBM part in learning the ground state. 
    This is helpful to reduce the number of variational parameters and helps alleviate the difficulty of optimizing a huge number of parameters. Especially, the wave function nodes and quantum-entanglement structure can be preconditioned. Otherwise, a complicated RBM complex parameter set has to be prepared (Fig.~\ref{Fig:RBM_PP}).
\end{enumerate}

In Ref.~\citen{Nomura_2017}, benchmark calculations of the RBM+PP wave function are performed using the Hubbard model on an $8\times8$ square lattice at half filling. 
Half filling is a special filling where the quantum Monte Carlo method can avoid the sign problem; therefore, the reference data for the ground state energy is obtained from the quantum Monte Carlo method.

\begin{figure}[t]
\begin{center}
\includegraphics[width=0.49\textwidth]{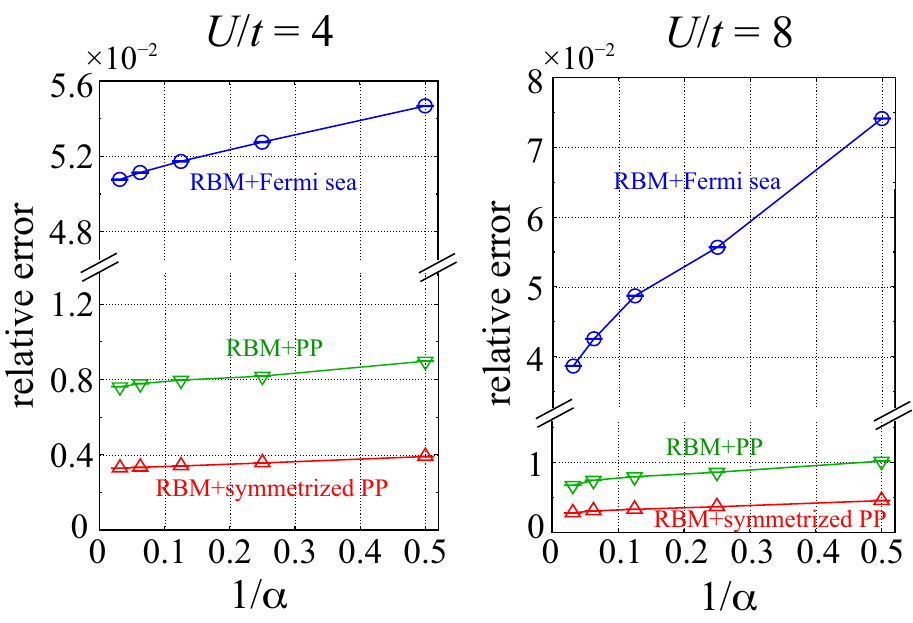}
\caption{
Accuracy of the variational energy of RBM wave function for two-dimensional Hubbard model on an $8\times8$ square lattice at half filling. 
$\alpha = M/2N$ specifies the size of the RBM.
As a comparison, the RBM is combined with the Fermi-sea state having no variational parameters (blue), the PP state (green), and the symmetrized PP state (red). 
Adapted with permission from Ref.~\citen{Nomura_2017}. Copyright 2017 by the American Physical Society.
}
\label{Fig:Benchmark_Hubbard}
\end{center}
\end{figure}

Figure~\ref{Fig:Benchmark_Hubbard} shows the comparison of accuracy among the different combinations; the RBM is combined with the Fermi-sea state (blue), the PP state (green), and the symmetrized PP state (red).
In the order of the Fermi-sea state, the PP state, and the symmetrized PP state, the combined wave function can increasingly incorporate sophisticated quantum correlations by small numbers of hidden variables $M$. 
If one took the simple product state in real space as in the case of Ref.~\citen{Carleo_2017}, the accuracy would be much worse than the case of the simple Fermi sea. 
Thus, the combination leads to a drastic improvement in the accuracy of the energy. 
We could also improve the energy by brute-force increasing the number of hidden units in the RBM part, however, we see that combining with a sophisticated wave function is a more efficient way to improve accuracy. 
We note that Fig.~\ref{Fig:Benchmark_Hubbard} was obtained by allowing only real and translationally invariant hidden parameters of RBM. 
If these constraints would be relaxed, the accuracy is expected to increase~\cite{Nomura_2021_JPCM,Reh_2023}.

Practical optimization of artificial neural networks often suffers from tradeoffs. In principle, increasing the number of parameters improves expressivity, but in practice, this makes numerical optimization harder and harder.
Therefore, the combination with the PP state and the resulting reduction of the variational parameters to achieve a certain level of accuracy is useful when considering applications to large system sizes, which is necessary to characterize the system in the thermodynamic limit.
Indeed, one of the first fruitful applications beyond benchmarks has been achieved by the combination of RBM and PP~\cite{Nomura_2021_PRX} (see Sec.~\ref{Sec:application} for more detail).

At the same time, there are ongoing attempts to improve optimization methods to stably optimize large deep neural networks.
Of note is an efficient SR optimization scheme when the number of parameters is much larger than Monte Carlo sampling, named minSR method~\cite{Chen_2024}.
Using the minSR method and optimizing as many as ${\mathcal O}(10^{5-6})$ variational parameters, Ref.~\citen{Rende_2024} using vision transformer and Ref.~\citen{Chen_2024} using residual neural network have achieved even better accuracy than the RBM+PP wave function with ${\mathcal O}(10^4)$ parameters for two-dimensional (2D) $J_1$-$J_2$ Heisenberg model in a frustrated regime $J_2/J_1=0.5$ (see Sec.~\ref{sec:J1J2} for the discussion of the phase diagram of this model).

\subsubsection{Other extensions}

So far, we have focused on describing the progress in the ground-state calculations. 
However, the variational method based on artificial neural networks can be extended to a wide variety of other applications such as
\begin{itemize}
    \item excited-state calculations~\cite{Choo_2018,Hendry_2019,Nomura_2020,Nomura_2021_JPCM,Vieijra_2020,Yoshioka_2021}
    \item real-time dynamics~\cite{Carleo_2017,Czischek_2018,Schmitt_2020}
    \item open quantum systems~\cite{Nagy_2019,Hartmann_2019,Vincentini_2019,Yoshioka_2019}
    \item finite-temperature calculations~\cite{Irikura_2020,Nomura_2021_PRL}
\end{itemize}
Please refer to the references for more details.

\section{Applications}
\label{Sec:application}

As discussed in Sec.~\ref{Sec:VMC_NN}, since the introduction of RBM variational ansatz in 2017~\cite{Carleo_2017}, there have been various attempts and benchmarking efforts to make the neural-network-based VMC approach a truly useful quantum many-body solver. 
Recently, these efforts have begun to bear fruit in generating productive runs for the analysis of many-electron systems.
Here, we present several examples.

\subsection{Fundamental many-body Hamiltonians}

\subsubsection{$J_1$-$J_2$ Heisenberg model on square lattice}
\label{sec:J1J2}

\begin{figure}[t]
\begin{center}
\includegraphics[width=0.49\textwidth]{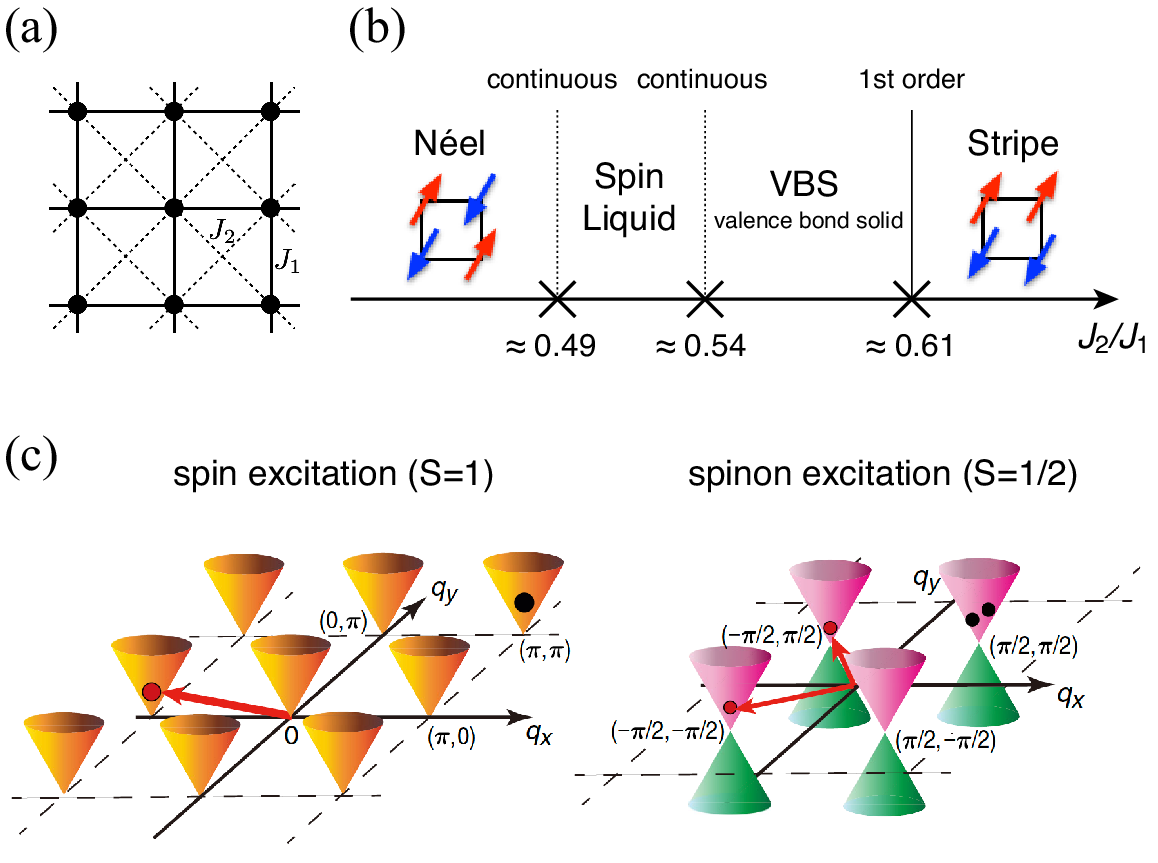}
\caption{
(a) Interactions in the $J_1$-$J_2$ Heisenberg model on the 2D square lattice and (b) ground-state phase diagram obtained using the RBM+PP wave function.
(c) Schematic figure for dispersions of spin and spinon excitations in the QSL phase. $S\!=\!1$ spin excitations constitute at least two $S\!=\!1/2$ spinon excitations. 
Adapted from Ref.~\citen{Nomura_2021_PRX}. CC BY 4.0.
}
\label{Fig:J1J2}
\end{center}
\end{figure}

We start the discussion from $J_1$-$J_2$ Heisenberg model on a 2D square lattice, which is also known as an effective model for Mott insulating cuprate parent compounds. 
The Hamiltonian is given by 
\begin{eqnarray}
\mathcal H =  J_1  \sum_{ \langle i,j \rangle  } {\bm S}_i \cdot {\bm S}_j  + J_2  \sum_{ \langle \langle i,j \rangle \rangle  } {\bm S}_i \cdot {\bm S}_j, 
\end{eqnarray}
where $J_1$ ($J_2$) is the nearest neighbor (next nearest neighbor) spin-spin interaction [Fig.~\ref{Fig:J1J2}(a)]. 
We consider a situation where both interactions are antiferromagnetic ($J_1,\ J_2>0$).

In this Hamiltonian, $J_1$ interaction favors N\'eel-type spin configuration, whereas $J_2$ interaction favors stripe-type spin configuration. 
Therefore, they compete with each other by geometrical frustration. 
The frustration effect becomes strong around $J_2/J_1 \approx 0.5$, and pronounced quantum fluctuations may lead to the realization of an exotic quantum phase (QSL) in which the spin orientations are not aligned even at absolute zero temperature.

Numerical analysis of this model is quite hard because the negative sign problem prevents the application of unbiased quantum Monte Carlo methods and the quantum correlations in the possible QSL phase are highly nontrivial.
Due to this difficulty, the ground-state phase diagram had not been settled, despite extensive numerical effort~\cite{Jiang_2012,Hu_2013,Gong_2014,Morita_2015,Wang_2016,Wang_2018,Haghshenas_2018,Liu_2018}.

In Ref.~\citen{Nomura_2021_PRX}, the RBM+PP wave function is introduced to capture nontrivial quantum correlations. 
To apply the RBM+PP wave function to spin systems, the Gutzwiller factor is applied to prohibit double occupancy. 
On top of the state-of-the-art accuracy achieved by RBM+PP, uncertainties in the estimate of the thermodynamic limit are successfully resolved by cross-checking correlation ratio~\cite{Kaul_2015} and level spectroscopy~\cite{Nomura_1995}. 
These two analyses utilize the properties in the ground and excited states, respectively.
Since there is a close one-to-one relationship between ground and excited states, the agreement between the two numerical analyses greatly improves the reliability of the results.

Figure~\ref{Fig:J1J2}(b) shows the obtained phase diagram, confirming the existence of QSL phase in a finite $J_2/J_1$ region ($0.49 \! \lesssim \! J_2/J_1  \! \lesssim  \! 0.54$).  
Investigation of excitation structure to characterize the obtained QSL reveals an emergent property arising from the fractionalization of the spin into spinon. 
Namely, experimentally observable spin excitations are interpreted as the composite excitation of at least two hidden spinons. In other words, Dirac-type linear {\it spinon} dispersion generates Dirac-type gapless {\it spin} excitation at high-symmetry momenta [Fig.~\ref{Fig:J1J2}(c)]. 
This property is consistent with $Z_2$ nodal spin liquid in the characterization of QSL~\cite{Wen_2002}.

Attempts to achieve more accurate variational computation are not limited to artificial neural networks but also open to other variational ans\"{a}tze such as tensor networks.
In the case of the 2D $J_1$-$J_2$ Heisenberg model, the analyses using DMRG~\cite{Wang_2018}, VMC~\cite{Ferrari_2020}, and PEPS (projected entangled-pair states)~\cite{Liu_2022} gave qualitatively consistent results with the RBM+PP results~\cite{Nomura_2021_PRX} in the sense that QSL and valence bond solid phases exist between N\'eel and stripe AF phases (we note that there also exists a recent study which states the absence of QSL phase~\cite{Qian_2024}). 
Since different variational ans\"{a}tze are complementary to each other, cross-checking between various approaches can also improve the reliability of the result.

\subsubsection{Heisenberg model on pyrochlore lattice}

Recently, the existence of the QSL on the 3D Heisenberg model on the pyrochlore structure
was reported by using the RBM+PP variational wave function.
The QSL is stabilized after the emergent dimensional reduction to 2D caused by the spontaneous symmetry breaking.
The nature of QSL shares common properties with the QSL in $J_1$-$J_2$ model reviewed in Sec.~\ref{sec:J1J2} in the sense that the excitation has gapless points and the fractionalization into spinons is representable by the quasiparticle in a projected nodal superconductor.
Readers are referred to Ref.~\citen{Pohle_arXiv} for details 
(see also a study~\cite{Astrakhantsev_2021} that partly uses neural-network quantum states).

\subsection{Ab initio low-energy Hamiltonians}
\label{sec:ab_initio}

So far, applications of the machine learning as the quantum many-body solver are mostly limited to simplified theoretical toy models such as the Heisenberg and Hubbard models. 
On the other hand, utilizing RBM to understand materials properties quantitatively on the first principles level, for instance, by using MACE, has also started generating fruitful results and has shown the power of the neural network even when the Hamiltonians get complicated. 
In this subsection, we review such achievements in the two examples, first the studies on the QSLs
in molecular solids, called dmit salts, and next, studies on SC in copper-oxide high-$T_c$ superconductors. 
Both of them have succeeded in reproducing the material dependence quantitatively and in extracting the universal properties with insights into the mechanism common in different compounds.

\subsubsection{dmit salts}

Molecular solids (organic conductors) host various electronic phases such as SC and AF Mott insulating states. Especially, $\beta'$-$X$[Pd(dmit)$_2$]$_2$ with $X$=EtMe$_3$Sb has quasi-2D anisotropic triangular lattice structure under large geometrical frustration effects and shows QSL properties in the Mott insulator phase. 
However, the nature of the QSL was largely unknown. 
Recent {\it ab initio} work~\cite{Ido_2022} successfully explored five compounds comprehensively with different $X$ using 2D {\it ab initio} Hamiltonians with the help of the machine learning. 

Despite a large number of atoms in the unit cell of such molecular-solid compounds, the band structure near the Fermi level is surprisingly simple, consisting of only LUMO (lowest unoccupied molecular orbital) and HOMO (highest occupied molecular orbital). 
The sparse bands near the Fermi level assure the strong correlation ascribed to the poor mutual screening. 
In addition, large inter-molecular distance leads to narrow bandwidths. 
Both cooperatively lead to strongly correlated electron systems with various types of Mott insulators. 

\begin{figure}[t]
  \begin{center}
    \includegraphics[width=0.8\linewidth]{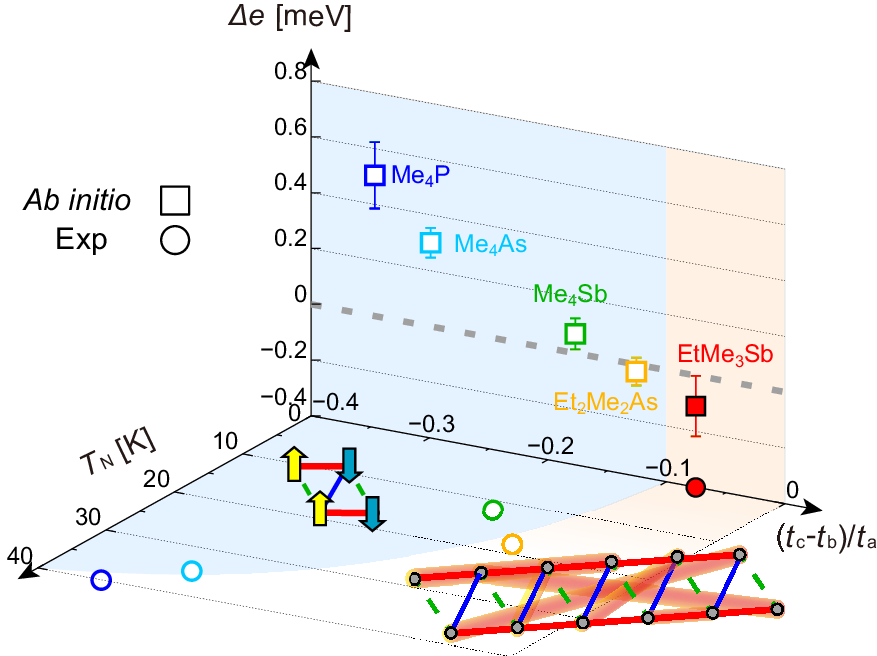}
  \caption{
    Ground-state phase diagram revealed by $ab$ $initio$ simulations for dmit compounds.
    Stability of electronic phases for {\it ab initio} Hamiltonians, plotted for $X$=Me$_4$P, Me$_4$As, Me$_4$Sb, Et$_2$Me$_2$As, and EtMe$_3$Sb aligned from the left (small $(t_c-t_b)/t_a$) to the right, where $t_a, t_b, t_c$ are the {\it ab initio} hopping parameters along the three directions on the anisotropic triangular lattice. 
    Vertical plane: Square symbols show the energy difference $\Delta e=e_{\rm QSL}-e_{\rm AF}$ per site between the energy of the QSL and that of the AF phase calculated for $16 \times 16$ lattice, where the energies of two lowest states, the QSL and collinear AF states, are compared. 
    Negative $\Delta e$ indicates that the QSL is the ground state. 
    The results with error bars were obtained after the extrapolation of the energy variance to zero by using the RBM and the first Lanczos methods.
    Bottom plane: Circles show the material dependence of the N\'{e}el temperature $T_N$ observed in the experiments~\cite{Nakamura_2001,Fujiyama_2019}.
    For the both planes, open symbols show the AF states and the filled ones are the QSL. Blue and orange shades represent the region of the collinear AF and the QSL phases, respectively. 
    Reprinted from Ref.~\citen{Ido_2022}. CC BY 4.0.
    }
\label{fig:dmit_phase_diagram}
\end{center}
\end{figure}

{\it Ab initio} electronic Hamiltonians without any adjustable parameters were derived for the half-filled HOMO band of five dmit compounds by using MACE based on the experimental structure without optimizing lattice structures except for positions of hydrogen atoms~\cite{Nakamura_2012,Yoshimi_2021}. 
These {\it ab initio} Hamiltonians were solved by the VMC combined with the RBM. 
The RBM together with the first Lanczos step is particularly useful for the accurate estimate of the energy competition between the QSL and the AF states, which enables to correctly reproduce the overall experimental phase diagram at low temperatures of 
$\beta'$-$X$[Pd(dmit)$_2$]$_2$. 
$X$=Me$_4$P, Me$_4$As, Me$_4$Sb, Et$_2$Me$_2$As exhibit the AF phases and $X$=EtMe$_3$Sb exhibits the QSL phase in agreement with the experimental phase diagram as is shown in Fig.~\ref{fig:dmit_phase_diagram}. 
The QSL for $X$=EtMe$_3$Sb shows 1D nature characterized by algebraic decay of spin correlation along one direction, while exponential decay in the other direction, indicating dimensional reduction from 2D to 1D.  
The 1D nature indeed accounts for the experimental specific heat, thermal conductivity, and magnetic susceptibility. 
The identified QSL, however, preserves 2D nature as well consistently with spin fractionalization into spinon with Dirac-like gapless excitations and reveals duality bridging the 1D and 2D QSLs.
The Dirac-like gapless excitation is common to the $J_1$-$J_2$ model discussed above, implying a universal nature of the QSL.


\begin{figure}[t]
  \begin{center}
    \includegraphics[width=0.8\linewidth]{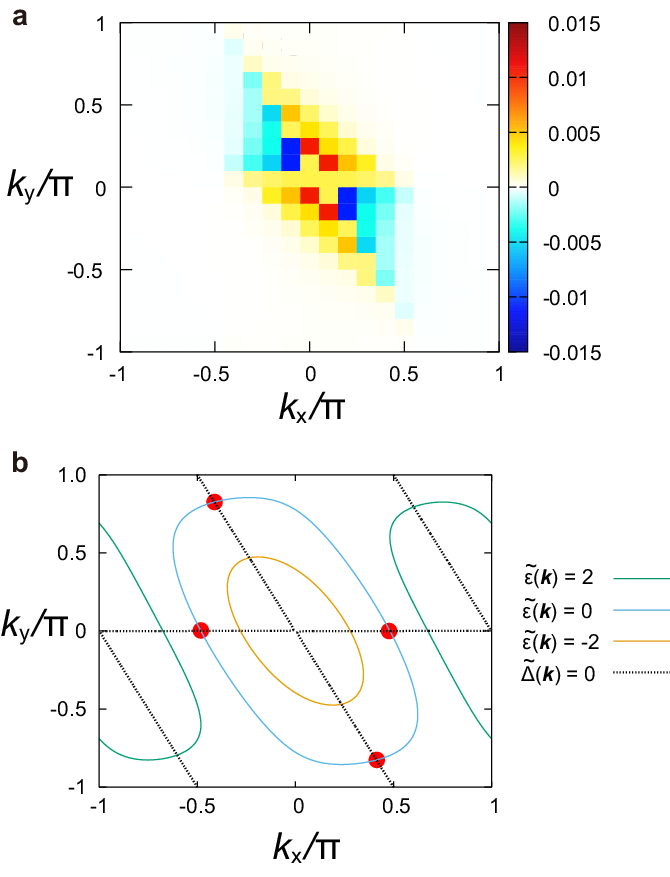}
  \caption{
    Dispersion of excitation inferred from the optimized variational wave function. $f(\bm {k})$ shown in {\bf a} is obtained from the Fourier transform of the optimized $f_{ij}$ which clearly shows sign change with two nodal lines. This structure is nearly perfectly fit by that of the $d$-wave BCS mean-field wave function given by Eq.~(\ref{BCSfij}) and in {\bf b}. 
    The intersections of the variational ``Fermi line" $\tilde{\epsilon}(\bm{k})=0$ (blue curve) and the vanishing gap line $\tilde{\Delta}(\bm{k})=0$ (dotted black lines), indicated by the red circles in {\bf b}, are the estimated gapless nodal points.
    Reprinted from Ref.~\citen{Ido_2022}. CC BY 4.0.
    }
\label{fitting}
\end{center}
\end{figure}

The gapless nature of the spin excitations is clear from the power law decay of the spin correlation and is consistent with the structure of the RBM+PP wave function for the PP part. Figure~\ref{fitting} demonstrates that the structure of $f_{ij}$ in Eq.~(\ref{eq_PP_state}) after the Fourier transform is well fitted by that obtained for the $d$-wave SC wave function in the BCS mean-field approximation given by
\begin{eqnarray}
  f(\bm{k})  =  \frac{\tilde{\Delta}(\bm{k})}{\tilde{\epsilon}(\bm{k})+\sqrt{\tilde{\epsilon}(\bm{k})^2+\tilde{\Delta}(\bm{k})^2}},
\label{BCSfij}
\end{eqnarray} 
where $\tilde{\epsilon} (\bm{k})$ and $\tilde{\Delta} (\bm{k})$ are the bare dispersion and the gap function of the BCS Hamiltonian, respectively.
This implies that the Dirac-like spinon excitation is equivalent to the nodal quasiparticle excitation of a $d$-wave superconductor. 
The gapless nature is retained even after the projection to the Mott insulator because the two-spinon composite excitation (corresponding to a hidden gapless particle-hole pair excitation of the quasiparticles at the nodal momenta in the $d$-wave SC wave function projected to the insulator) generates one observable spin excitation, which does not destroy the Mott insulating nature. 

\subsubsection{Cuprates}

The cuprate superconductors have commonly layered perovskite structure and belong to typical strongly correlated quantum materials, where the conventional DFT fails in reproducing the Mott insulating phase in the undoped parent compounds. 
The SC emerges upon carrier doping into AF Mott insulators.
In the cuprates, the $d$-wave SC state is severely competing with other orders, such as spin and charge stripes (periodic alignment of spin and charge in stripe configurations in real space) or AF states, and the observed superconducting transition temperature $T_{c}$ sensitively depends on the materials widely ranging from above $130\, \text{K}$ to below $10\, \text{K}$. 

Understanding and reproducing these diverse phenomena without relying on adjustable parameters as well as identifying the universal parameter to control the strength of the SC in the cuprates are the major challenges in condensed matter physics since the discovery in 1986~\cite{Bednorz_1986}, while the {\it ab initio} calculations had been faced with many difficulties.
When {\it ab initio} calculations are able to reproduce systematic materials dependence quantitatively by solely relying on their crystal structures, they provide us with valuable insights into the universal mechanism behind and into the principal components for the enhancement of SC beyond existing materials.

The effective Hamiltonians were derived in the form of Eq.~(\ref{eq:Hubbard})
by using the state-of-the-art MACE using cGW-SIC+LRFB for four compounds, CaCuO$_2$ ($T_{c}^{\rm opt} \! \sim \!  110$ K), Bi$_2$Sr$_2$CuO$_6$ (Bi2201) ($T_{c}^{\rm opt} \! \sim \!  10$-$40$ K), Bi$_2$Sr$_2$CaCu$_2$O$_8$ (Bi2212) ($T_{c}^{\rm opt} \! \sim \! 85$-$100$ K), and  HgBa$_2$CuO$_4$ (Hg1201) ($T_{c}^{\rm opt} \! \sim \! 90$ K)~\cite{Moree_2022,Hirayama_2019},
where $T_{c}^{\rm opt}$ is $T_c$ at the optimum hole doping reported in experiments.
The derived Hamiltonian parameters are rather similar among the four materials. 
For instance, the ratio between the onsite interaction $U$ and the nearest neighbor hopping $t_1$ defined by $U/|t_1|$ lies in a narrow range between 7.3 and 10.4. 
Then, the question is whether one can reproduce the diversity in the material dependence of physical properties such as $T_c$ from the similar Hamiltonians.

The Hamiltonians were solved by using the VMC using the RBM+PP type wave function, where the interaction and hopping were considered up to the 9th neighbor on the square lattice in the calculations~\cite{Schmid_2023}. 
The variational wave function was further improved~\cite{Misawa_2019} by the doublon-holon correlation factor and the first Lanczos step. 
They were also used to determine the lower energy state in the severe competition of the SC, AF, and stripe states. The advantage of the RBM is that such additional procedures can easily be combined to improve the accuracy.

\begin{figure}[tb]
\begin{center}
\includegraphics[width=0.8\linewidth]{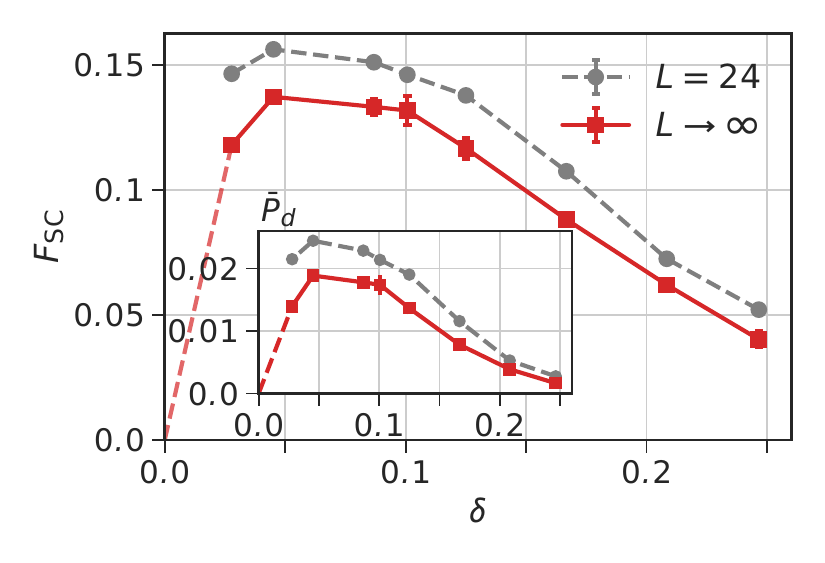} 
\caption{The SC order parameter $F_{\rm SC}$ as a function of hole doping $\delta$ for CaCuO$_2$. 
The gray filled circles show the values of $F_{\rm SC}$ at $24 \times 24$ square lattice, while the red squares are the size extrapolated values (thermodynamic limit) of $F_{\rm SC}$. 
The inset shows the corresponding SC correlation function $\bar{P}_d \equiv F_{\rm SC}^2$ at the converged longest distance.
Reprinted from Ref.~\citen{Schmid_2023}. CC BY 4.0.
}
\label{fig:Fsc_delta}
\end{center}
\end{figure}

The VMC combined with the RBM first confirmed the dominance of the $d$-wave SC over severely competing AF and the charge/spin stripe states in all the materials studied, where the SC states have around ten meV lower energy. 
The dominance of the SC consistently with the experiments is noticeable because the simple Hubbard model with only $U$ and $t_1$ is governed by the stripe phase and the charge uniform SC state is an excited state~\cite{Zhao_2017,Zheng_2017,Ido_2018,Darmawan_2018}. 
The reason for the reversal of the stability is mainly due to the effect of the offsite interaction $V_{ij}$, which severely suppresses the stripe phases~\cite{Ohgoe_2020}. 
The SC order parameter $F_{\rm SC}$ shows a dome-like structure as a function of the carrier density $\delta$ as is plotted in Fig.~\ref{fig:Fsc_delta}, which is quantitatively consistent with the experimental results.

\begin{figure}[t]
    \begin{center}
    \includegraphics[width=0.8\linewidth]{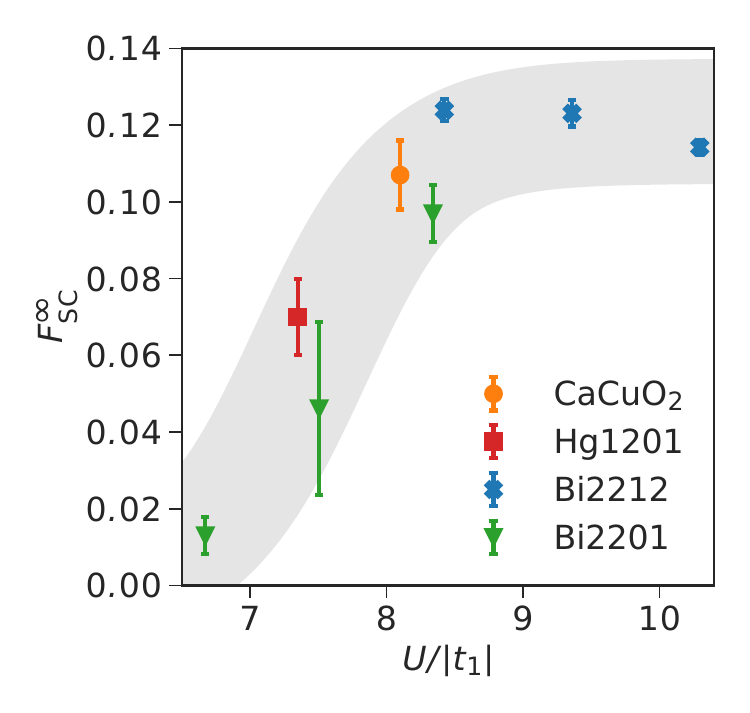}
    \caption{$F_{\rm{SC}}^{\infty}$ ($F_{\rm{SC}}$ in the thermodynamic limit) as a function of $U/|t_1|$ for the four cuprate compounds at $\delta = 0.167$. 
    Reprinted from Ref.~\citen{Schmid_2023}. CC BY 4.0.
    }    
    \label{fig:FSC_vs_U/t}
    \end{center}
\end{figure}

The material dependence of the SC order parameter $F_{\rm SC}$ plotted in Fig.~\ref{fig:FSC_vs_U/t} indicates diversity despite the similar Hamiltonian parameters. 
In fact, $F_{\rm SC}$ has a systematic and sensitive dependence on $U/|t_1|$, demonstrating that $U/|t_1|$ is the principal component to control the SC. 
Most of the materials are located in the weak coupling side below the optimum of $U/|t_1|\sim 9$.

\begin{figure}[t]
    \begin{center}
    \includegraphics[width=0.4\textwidth]{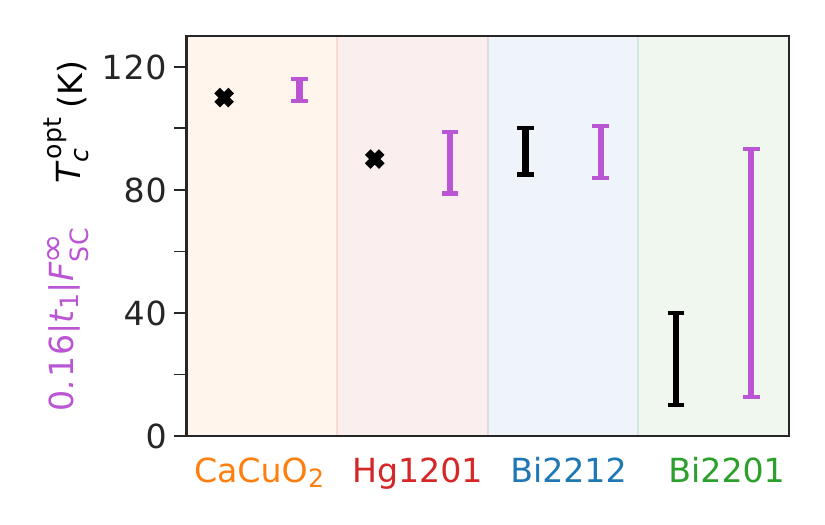}
    \caption{Experimental $T_{c}^{\rm{opt}}$ (black crosses or bars) in comparison to the $ T_{c}  = 0.16  \lvert t_1 \rvert  F_{\rm{SC}}^{\infty}$ scaling for each compound (purple bars). 
    Reprinted from Ref.~\citen{Schmid_2023}. CC BY 4.0.
    }
    \label{fig:TcComparisionExperiment}
    \end{center}    
\end{figure}

The obtained order parameter can be used to discuss the optimum $T_c$. 
It has turned out that a universal scaling law for the optimum $T_c^{\rm opt} =0.16|t_1|F_{\rm SC}$ is well consistent with the experiments for the four materials as is shown in Fig.~\ref{fig:TcComparisionExperiment}. 

\begin{figure}[t]
    \begin{center}
    \includegraphics[width=\linewidth]{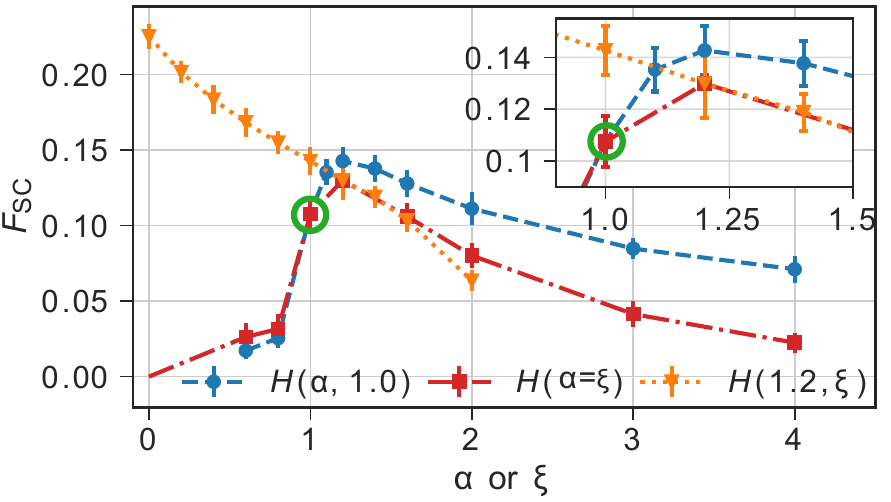} 
    \caption{Trend of SC order beyond the so far available materials. $F_{\rm SC}$ is plotted as functions of $\alpha$ or $\xi$, which are $U/\lvert t_1 \rvert$ or $V_{ij}/U$ scaled by the {\it ab initio} values, respectively, in the Hamiltonian $H(\alpha,\xi)$.  The results are for the $24 \times 24$ square lattice at $\delta = 0.167$ hole doping.
    The inset is a zoom-in plot around the peak $0.9 < \alpha, \ \xi < 1.5$. The green circles indicate the {\it ab initio} points for carrier doped CaCuO$_2$ at the optimum doping.
    Adapted from Ref.~\citen{Schmid_2023}. CC BY 4.0.}
    \label{fig:CaCuOPdScalingZoom}
    \end{center}
\end{figure}

The present method can also monitor the effect of interaction by the parameter search beyond the {\it ab initio} Hamiltonian to gain insights into the strategy for the materials design. Beyond the available materials, the SC order parameter can be optimized by tuning the onsite interaction to the peak in Fig.~\ref{fig:CaCuOPdScalingZoom} (blue and red curves) and by reducing the offsite interaction strength $V_{ij}$ (orange curve) provided that the stripe order does not intervene.

\section{Concluding remarks}
\label{sec:conclusion}

In this review, we have seen that machine learning methods can be a powerful tool for analyzing quantum many-body problems. 
As concluding remarks, we raise several points towards further growth of the field.

\begin{itemize}
    \item {\bf Further methodological development}:
    Productive applications using artificial neural networks to date have often employed the combination with physics-inspired ansatz such as the PP state (see Sec.~\ref{Sec:application}). 
    One of the reasons for this is to avoid introducing a huge number of variational parameters and to avoid optimization troubles.
    Without the combination, artificial neural networks would require a huge amount of variational parameters to reach state-of-the-art accuracy. For example, ten times more parameters than the RBM+PP case are needed for the transformer-based~\cite{Rende_2024} and residual-neural-network-based~\cite{Chen_2024} wave function to beat the accuracy of RBM+PP wave function~\cite{Nomura_2021_PRX} for the 2D $J_1$-$J_2$ Heisenberg model (see Sec.~\ref{sec:extension_challenging}). 
    It is certainly necessary to pursue more powerful optimization methods (the development of minSR~\cite{Chen_2024} is along this direction) and to consider appropriate network designs that allow quantum states to be represented by more compact neural networks.
    \item {\bf Novel algorithm}:
    So far, the hidden variables in the artificial neural network developed for the quantum man-body solver are mostly taken to be classical variables such as Ising spins. To incorporate the quantum entanglement more efficiently, there exist proposals to replace the classical hidden variables with hidden fermions~\cite{Moreno_2022,Imada_2024}. 
    It is intriguing to further develop algorithm along this line. It is expected to be useful as well to understand the structure of the fractionalization in strongly correlated electrons. 
    \item {\bf White-boxing}: 
    The process of machine learning is often described as a black box. 
    What the artificial neural network learned through optimization is extremely nontrivial. 
    Rather than treating machine learning methods as black-box tools that provide useful results, the challenge for the future is how to extract essential physical insights from them.
    \item {\bf Toward a truly useful tool}:
    In this review, we have provided several examples of productive applications to unravel the physics of difficult problems.
    However, many researches are still limited to benchmarking.
    More productive applications are certainly necessary for the machine learning method to be recognized as a truly useful tool.
    \item {\bf Cross-check}:
    As discussed in Sec.~\ref{sec:J1J2}, the agreement of results obtained by different types of variational wave functions has solidified the conclusion on QSL in the case of the 2D $J_1$-$J_2$ Heisenberg model.
    Such cross-checks will become increasingly important in settling controversies in unsolved quantum many-body problems.
    To make comparisons among different numerical algorithms meaningful, a consistent accuracy metric is desired.
    With this in mind, attempts are being made to build a platform for comparing accuracy among different algorithms and different Hamiltonians~\cite{Wu_arXiv}.
    Further development of variational approaches, including both classical and quantum algorithms, and reliable comparisons between methods will certainly contribute to a deeper understanding of quantum many-body physics.
\end{itemize}

\begin{acknowledgment}
\noindent
{\it Acknowledgments.}
The authors acknowledge useful discussions and collaborations with 
G. Carleo, A. S. Darmawan, M. Hirayama, K. Ido,  T. Misawa, J.-B. Mor\'{e}e, F. Nori, M.T. Schmid, Y. Yamaji, K. Yoshimi, and  N. Yoshioka.  
Y.N. thanks C. Falter for comments. 
Y.N. and M.I. acknowledge support from MEXT as ``Program for Promoting Researches on the Supercomputer Fugaku'' (Grant No. JPMXP1020230411).
Y.N. is supported by Grant-in-Aids for Scientific Research (JSPS KAKENHI) (Grant Nos. JP23H04869, JP23H04519, and JP23K03307) and JST (Grant No. JPMJPF2221).
M.I. is financially supported by MEXT KAKENHI, Grant-in-Aid for Transformative Research Area (Grant Nos. JP22H05111 and JP22H05114). 
\end{acknowledgment}

\profile{Yusuke Nomura}{was born in Tokyo, Japan, in 1988. 
He received Ph.D. in 2015 from University of Tokyo and his thesis has been published from Springer as an outstanding thesis. 
After 
a postdoctoral fellow at \'Ecole Polytechnique (2015-2016), 
a research associate at University of Tokyo (2016-2019), 
a research scientist at RIKEN (2019-2022), 
and
an associate professor at Keio University (2022-2024), 
he is now a professor at Institute for Materials Research, Tohoku University. 
His major is quantum many-body physics and computational materials science. 
Recently, a Japanese textbook on neural-network quantum states, co-authored with N. Yoshioka, was published.
}
\profile{Masatoshi Imada}{was born in Kumamoto Prefecture, Japan, in 1953. He received Ph.D. in 1981 from University of Tokyo. He was a research associate at
University of Tokyo (1981-1986), a lecturer and an associate professor at Saitama University (1986-1990), an associate professor (1990-1997), and a professor (1997-2019) at University of Tokyo. He was a fellow at Toyota Physical and Chemical Research Institute and a research professor at Waseda University (2019-2023), a visiting professor at Sophia University (2018-), as well as a professor emeritus at University of Tokyo (2019-). He has been working on condensed matter theory, computational physics, and statistical physics. The main research subjects are strongly correlated quantum systems including metal-insulator transitions,
superconductivity, magnetism, and quantum phase transitions. His research also covers methodology for quantum many-body problems.}

\bibliographystyle{jpsj}
\bibliography{main}


\end{document}